\documentstyle[aps,12pt,epsfig,amssymb]{revtex}

\newcommand{\bm}{\bibitem}

\renewcommand{\vec}[1]{\mbox{\boldmath$#1$\unboldmath}}
\newcommand{\cp}{\chi^{(+)}}
\newcommand{\cm}{\chi^{(-)*}}

\newcommand{\vv}{V_{cn}({\bf r}_1)}
\newcommand{\ri}{{\bf r}_i}
\newcommand{\ro}{{\bf r}_1}
\newcommand{\ak}{{\bf k}_a}
\newcommand{\cq}{{\bf k}_c}

\newcommand{\rn}{{\bf r}_n}
\newcommand{\nq}{{\bf k}_n}
\newcommand{\we}{\Psi^{(+)}_a(\xi_a,{\bf r}_1,{\bf r}_i)}

\newcommand{\fa}{ _2F_1(1-i\eta_a,1-i\eta_c;2;D(0))}
\newcommand{\fB}{ _2F_1(-i\eta_a,-i\eta_c;1;D(0))}
\tighten
\begin{document}
\draft
\title{Study of Postacceleration Effects in  the Coulomb Dissociation of
Neutron Halo Nuclei}
\date{\today}
\author{Prabir Banerjee$^{1}$, Gerhard Baur$^{2}$, Kai Hencken$^{3}$, 
Radhey Shyam$^{4}$, and Dirk Trautmann$^{3}$}
\address{ 
$^1$
Physics Group, Variable Energy Cyclotron Centre, Calcutta-700064, India\\
$^2$
Institut f\"ur Kernphysik (Theorie), Forschungszentrum J\"ulich,
D-52425 J\"ulich, Germany\\
$^3$ 
Institut f\"ur Physik, Universit\"at Basel, Klingelbergstr. 82,
D-4056 Basel, Switzerland\\
$^4$
Saha Institute of Nuclear Physics, Calcutta 700064, India}
\maketitle

\begin{abstract}
We study the breakup of one-neutron halo nuclei in the Coulomb field of
a target nucleus. In the post-form distorted wave Born approximation theory
of this reaction, with only Coulomb distortions in the entrance and the
outgoing channels, an analytic solution for the breakup $T$-matrix is known.
We study  this $T$-matrix and the corresponding cross-sections numerically.
This formula can be related to the first order semiclassical 
treatment of the electromagnetic dissociation. This theory contains 
the electromagnetic interaction between the core and the target nucleus 
to all orders. We show that higher order effects (including 
postacceleration) are small in the case of higher beam energies and 
forward scattering. We investigate the beam energy dependence of the
postacceleration effects. They are found to be quite important for
smaller beam energies (slow collisions), but almost negligible 
at larger ones.
\end{abstract}
\newpage
\section{Introduction}
Breakup processes, in nucleus-nucleus collisions, are complicated, in whatever
way they are studied.  Coulomb breakup (CB) is a significant reaction
channel in the scattering of halo nuclei from a heavy target nucleus
(see, e.g., \cite{han95,tani96,bau00,shy01}). With the operation of
exotic beam facilities all over the world, these reactions (previously
restricted essentially to deuteron induced reactions) have come into sharp
focus again. CB provides a convenient way to put constraints on the 
structure of these nuclei \cite{ber88,bes91}. This is of interest  
also for nuclear astrophysics, since the breakup cross 
section can be related to the photo-dissociation cross section and to
radiative capture reactions relevant for nuclear astrophysics \cite{BaurR96}.

The breakup reactions of the halo nuclei have been investigated 
theoretically by several authors, using a number of different approaches
(see, e.g., \cite{bau00,cha00}, for an extensive list of references). 
One often used method has been to treat the breakup reaction as the inelastic 
excitation of the projectile from its ground state to the continuum
\cite{ber91}. The corresponding $T$-matrix is written in terms of
the prior form distorted wave Born approximation (DWBA) \cite{bau84}.
For pure Coulomb breakup,
the semiclassical approximation of this theory is the first order
perturbative Alder-Winther theory of Coulomb excitation \cite{ald75}.
It has recently been used to analyze the 
data on the breakup reactions induced by the one-neutron halo nuclei
$^{11}$Be \cite{nak94} and $^{19}$C \cite{nak99}. The question of the
higher order electromagnetic effects \cite{esb93,esb95,kid96,mel99,tos00}
was studied \cite{TypelB01} recently within this framework. They 
were found to be small, for both zero range as well as finite range wave
functions of the relative motion of the fragments within the projectile.
Also, in a dynamical description of the
breakup of these nuclei where the time evolution of the
projectile in coordinate space is described by solving the three-dimensional
time dependent Schr\"odinger equation treating the projectile-target
interaction as a time dependent external perturbation, the higher order
effects turned out \cite{ts01} to be only of the order of 10$\%$  
for beam energies in the range of 60 - 80 MeV/nucleon.

A direct breakup model (DBM) (which reduces to the Serber model 
in a particular limit \cite{bau00}) has been formulated within the
framework of the post form DWBA \cite{cha00,bau84}. An important advantage
of this model is that it can be solved analytically for the case of the
breakup of the neutron halo nuclei with the entrance and outgoing channels  
involving only the Coulomb distortions \cite{cha00,btr72,ShyamBB92}.
It constitutes an ideal ``theoretical laboratory" to investigate the physics
of the breakup reactions, its certain limiting cases, and its relation to
other models like the semiclassical approximation.  Particularly, the effect of
postacceleration (to be explained in more detail below) can be studied 
in a unique way within this approach.

The aim of this paper is to investigate the role of the higher order effects
(which includes postacceleration) in the Coulomb breakup of the 
one-neutron halo nuclei, $^{11}$Be and $^{19}$C, within the post form
DWBA theory of the breakup reactions. We calculate the triple and the double
differential cross sections of the fragments, observed in the breakup of 
these nuclei on a $^{208}$Pb target, within the
exact theory as well as within its first order approximation. We also 
calculate the relative energy distributions of the outgoing fragments
emitted in these reactions, within the two theories. Calculations have
been performed for a range of beam energies in order to investigate the
beam energy dependence of the higher order effects. In Sec. II, we present 
our theoretical model. The results of our calculations and their 
discussions are given in Sec. III.  Summary, conclusions, and the 
outlook of our work in given in Sec. IV.

\section{Description of the Theoretical Model}
\subsection{Coulomb wave Born approximation}
We consider the reaction $ a + t \rightarrow c + n + t $, where the
projectile $a$ breaks up into the charged fragment $c$ 
and the neutron $n$ in the Coulomb field of a target $t$.

The starting point of the Coulomb wave Born
approximation (CWBA) is the post form $T$ matrix of the breakup reaction
which is given by
\begin{eqnarray}
&&T = \int  d\xi d\ro d\ri \cm_c(\cq,{\bf r})\Phi^*_c(\xi_c)
\cm_n(\nq,{\bf \rn})\Phi^*_n(\xi_n)\vv \we.
\end{eqnarray}
The functions $\cm_c(\cq,{\bf r})$, and  $\cm_n(\nq,\rn)$ are
the distorted waves
for the relative motions of $c$ and $n$ with respect to $t$ and the
center of mass (c.m.)\ of the $c+t$ system, respectively. The arguments 
of these functions contain the corresponding Jacobi momenta and 
coordinates. $\Phi$'s are 
the internal state wave functions of the concerned particles
which depend on the internal coordinates $\xi$. The function $\we$
is the exact three-body scattering wave function of the projectile
with a wave vector $\ak$, which satisfies outgoing wave boundary
conditions.  The vectors $\cq$ and $\nq$ are the Jacobi wave
vectors of $c$ and $n$, respectively, in the final channel of
the reaction. The function $\vv$ represents the interaction
between $c$ and $n$. For the pure Coulomb breakup case, 
the function $\chi^{(-)}_c({\cq},{\bf r})$ is taken
as the Coulomb distorted wave (for a point Coulomb interaction between
the charged core c and the target) satisfying incoming wave
boundary conditions, and the function $\chi^{(-)}_n({\nq},{\rn})$ is
just a plane wave as there is no Coulomb interaction between the target
and the neutron.
The position vectors satisfy the following relations
\begin{eqnarray}
{\bf r} &=& \ri - \alpha\ro,~~ \alpha = {m_n\over m_a} \, ,   \\
{\bf r}_n &=& \gamma \ro +\delta \ri, ~~ \delta = \frac{m_t}{m_c+m_t},
~~  \gamma = (1 - \alpha\delta) \, .
\end{eqnarray}
In the distorted wave Born approximation, we write
\begin{eqnarray}
\we \approx \Phi_a(\xi_a,\ro)\cp_a(\ak,\ri).
\end{eqnarray}
In Eq.~(4), the dependence of $\Phi_a$ on ${\bf r}_1$
describes the relative motion of the fragments $c$ and $n$ in the ground
state of the projectile.  The function $\cp_a(\ak,\ri)$ is the Coulomb
distorted scattering wave describing the relative motion of the c.m.\ of
the projectile with respect to the target, satisfying outgoing wave boundary
conditions. 

The integration over the internal coordinates $\xi$, in Eq.~(1), gives
\begin{eqnarray}
\int d\xi\Phi^*_c(\xi_c)\Phi^*_n(\xi_n)\Phi_a(\xi_a,\ro)
& = & \sum_{\ell mj\mu} \langle \ell mj_n\mu_n|j\mu\rangle
 \langle j_c\mu_cj\mu|j_a\mu_a\rangle i^\ell \Phi_a(\ro),
\end{eqnarray}
with
\begin{eqnarray}
\Phi_a(\ro) & = & u_{\ell}(r_1) Y_{\ell m}({\hat{\bf r}}_1).
\end{eqnarray}
In Eq.~(6),  $\ell$  (the
orbital angular momentum for the relative motion between fragments
$c$ and $n$) is coupled to the spin of $n$ and
the resultant channel spin $j$ is coupled to spin
$j_c$ of the core $c$ to yield the spin of $a$ ($j_a$).
The $T$-matrix can now be written as
\begin{eqnarray}
T^{\textrm {CWBA}} & = & \sum_{\ell mj\mu}
\langle \ell mj_n\mu_n|j\mu\rangle
\langle j_c\mu_cj\mu|j_a\mu_a\rangle i^\ell
\hat{\ell}\beta_{\ell m}^{\textrm {CWBA}}(\cq,\nq;\ak),
\end{eqnarray}
where
\begin{eqnarray}
&&\hat{\ell}\beta_{\ell m}^{\textrm {CWBA}}(\cq,\nq;\ak)  = 
\int d\ro d\ri\cm_c(\cq,{\bf r})e^{-i\nq.\rn} \vv u_\ell (r_1) Y_{\ell m}
({\hat r}_1)\cp_a(\ak,\ri),
\end{eqnarray}
with ${\hat \ell} \equiv \sqrt{2\ell + 1}$.

Eq.~(8) involves a six dimensional integral which makes its 
computation quite complicated.  The problem gets further acute because
the integrand has a product of three scattering waves that exhibit an
oscillatory behavior asymptotically. Therefore, approximate methods have
been used, such as the zero range approximation (ZRA) in which we write,
$\vv\Phi_a(\ro) = D_0 \delta ({\bf r}_1)$ with $D_0$ being the zero range
normalization constant, (see, e.g., \cite{aus70,gle83,sat64}),
or the approximation used in \cite{btr72}, where the projectile
c.m.\ coordinate is replaced by that of the core-target
system (i.e. $\ri \approx {\bf r}$). Both these methods lead to a
factorization of the amplitude [Eq.~(8)] into two independent parts, which
reduces the computational complexity to a great extent. However, the
application of both these methods to the reactions of halo nuclei is
questionable (see, e.g., \cite{cha00}, for a detailed discussion).

In the finite range CWBA theory \cite{cha00}, the Coulomb
distorted wave of particle $c$, in the final channel, is written as 
\begin{eqnarray}
\chi^{(-)}_c(\cq,{\bf r}) & = & e^{-i\alpha{\bf K}.\ro}
                           \chi^{(-)}_c(\cq,\ri).
\end{eqnarray}
Eq.~(9) represents an exact Taylor series expansion about ${\bf r}_i$ if
$ {\bf K} = -i\nabla_{{\bf r}_i}$ is treated exactly. However, instead of
doing this we employ a local momentum approximation \cite{shyam85,braun74a},
where the magnitude of momentum ${\bf K}$ is taken to be
\begin{eqnarray}
K(R) & = &{\sqrt {{2m\over \hbar^2}(E - V(R))}}.
\end{eqnarray}
Here $m$ is the reduced mass of the $c-t$ system,
$E$ is the energy of particle $c$ relative to the target in the
c.m.\ system and $V(R)$ is the  Coulomb potential
between $c$ and the target $t$ separated by the distance $R$. Thus,
the magnitude of the momentum ${{\bf K}}$ is evaluated at
some separation $R$ which is held fixed for all the values of ${r}$.
For further details and the discussion on the validity of this
approximation, we refer to \cite{cha00,cha01}. 
 
On substituting Eq.~(9) into Eq.~(8), we obtain the following
factorized form of the amplitude $\beta^{\textrm{CWBA}}_{\ell m}$ 
\begin{eqnarray}
{\hat \ell}\beta^{\textrm{CWBA}}_{\ell m}(\cq,\nq;\ak) 
 & = & Z_{\ell m} 
\int d\ri \chi^{(-)*}_c(\cq,\ri) e^{-i\delta\nq.\ri}
\cp_a(\ak,\ri).
\end{eqnarray} 
where
\begin{eqnarray}
Z_{\ell m}& = & 
\int d\ro e^{-i{\bf k}_1.\ro}V_{cn}(\ro) u_\ell(r_1)
Y_{\ell m}({\hat r}_1), 
\end{eqnarray} 
where ${\bf k}_1 = \gamma\nq - \alpha {\bf K}$.

This amplitude differs from that of the ZRA studied earlier \cite{bau84}
as it allows the use of the full wave function for the relative motion 
of the fragments (corresponding to any value of $\ell$)
in the ground state of the projectile. However, it should be stressed
that as for as the postaccetelation effects are concerned, both the
amplitudes would lead to identical results.

The triple differential cross section of the reaction is given by
\begin{eqnarray}
{{d^3\sigma}\over{dE_cd\Omega_cd\Omega_n}} & = &
{2\pi\over{\hbar v_a}}\rho(E_c,\Omega_c,\Omega_n)
\sum_{\ell m}|\beta_{\ell m}^{\textrm {CWBA}}|^2,
\end{eqnarray}
where $\rho(E_c,\Omega_c,\Omega_n)$ is the appropriate three-body phase
space factor \cite{cha00}, and $v_a$ the velocity of particle $a$. 

On substituting the Coulomb distorted waves,
\begin{eqnarray}
\cm_c(\cq,{\ri}) & = &
 e^{-\pi\eta_c/2}\Gamma(1 + i\eta_c) e^{-i\cq.\ri}
 {_1F_1(-i\eta_c, 1, i(k_c r_i + \cq.\ri))} \,  , \\
                                          \nonumber \\
\cp_a(\ak,\ri) & = &
 e^{-\pi\eta_a/2}\Gamma(1 + i\eta_a) e^{i\ak.\ri}
 {_1F_1(-i\eta_a, 1, i(k_a r_i - \ak.\ri))}      \, ,
\end{eqnarray}
into Eqs.\ (11) and (13), one gets for the triple differential
cross section
\begin{eqnarray}
{{d^3\sigma}\over{dE_cd\Omega_cd\Omega_n}} = {32\pi^4\over{{\hbar}v_a}}
\rho(E_c,\Omega_c,\Omega_n)
{\eta_a\eta_c\over (e^{2\pi\eta_c}-1)(e^{2\pi\eta_a}-1)}|I|^2
\sum_{\ell} |Z_{\ell}^\prime|^2.
\end{eqnarray}
In Eqs.~(14--16), $\eta$'s are the Coulomb parameters for the
respective particles. In Eq.\ (16),
$I$ is the Bremsstrahlung integral \cite{nord} which can be
evaluated in the closed form:
\begin{eqnarray}
I &=& -i{\Big[}B(0){\Big(}{{dD}\over{d\Lambda}}{\Big)}_{\Lambda=0}
-\eta_a\eta_c)\fa \nonumber \\
& + & {\Big(}{{dB}\over{d\Lambda}}{\Big)}_{\Lambda=0} {\fB} {\Big]} \, ,
\end{eqnarray}
where
\begin{eqnarray}
B(\Lambda) = {4\pi\over{k^{2(i\eta_a+i\eta_c+1)}}}
{\Big[}(k^2 - 2{\bf k}.\ak -2\Lambda k_a)^{i\eta_a}
(k^2 - 2{\bf k}.\cq -2\Lambda k_c)^{i\eta_c}{\Big]},
\end{eqnarray}
\begin{eqnarray}
D(\Lambda) = {2k^2(k_ak_c+\ak.\cq)-4({\bf k}.\ak+\Lambda k_a)
({\bf k}.\cq+\Lambda k_c)\over
{(k^2 - 2{\bf k}.\ak -2\Lambda k_a)(k^2 - 2{\bf k}.\cq -2\Lambda k_c)}} \, ,
\end{eqnarray}
with
\begin{eqnarray}
{\bf k} &=& \ak - \cq -\delta\nq.
\end{eqnarray}
The factor $Z{^\prime}_{\ell}$ contains the projectile structure information
and is given by
\begin{eqnarray}
Z{^\prime}_{\ell} = \int dr_1 r^2_1 j_{\ell} (k_1 r_1)V_{cn}(r_1)
                     u_{\ell} (r_1).
\end{eqnarray}

It may be noted that the triple differential
cross sections with respect to relative and c.m.
coordinates of the fragments are related to those given by Eq.~(13) as
\begin{eqnarray}
{d^3\sigma}\over{dE_{c-n}d\Omega_{c-n}d\Omega_{t-(c+n)}} & = &
J\times {d^3\sigma}\over{dE_cd\Omega_cd\Omega_n}, 
\end{eqnarray}
where the form of Jacobian $J$ is the same as that given in \cite{fuc82}.
In Eq.~(14), $t-(c+n)$ corresponds to the coordinates of the relative motion
of the c.m. of the fragments $c$ and $n$ with respect to the target, while
$c-n$ corresponds to that of the relative motion between them. 

The CWBA description [Eqs.~(11-12)] includes the effects of
postacceleration, which refers to the situation where the core $c$ has
a larger final state energy than what one gets from sharing the kinetic
energy among the fragments according to their mass ratio. This effect
arises in  a purely classical picture \cite{BaurBK95} of the breakup process.
The nucleus $a=(c+n)$ moves up the Coulomb potential, loosing the
appropriate amount of kinetic energy. At an assumed ``breakup point'',
this kinetic energy (minus the binding energy) is supposed to be shared
among the fragments according to their mass ratio (assuming that the
velocities of $c$ and $n$ are equal). Running down the Coulomb barrier, the
charged particle $c$ alone (and not the neutron) gains back the Coulomb 
energy, resulting in its postacceleration. Of course this picture is based
on the purely classical interpretation of this process, and will be
modified in a quantal treatment, where such a ``breakup point'' does
not exist.  Postacceleration is clearly observed in the low energy deuteron
breakup, both in the theoretical calculations
as well as in the corresponding experiments (see, e.g., \cite{bau84,BaurT76}).
However, in the description of the Coulomb dissociation of halo nuclei at
high beam energies within this theory \cite{cha00,ShyamBB92,shyam93},
the postacceleration effects become negligibly small. We shall investigate
this point further for the $^{11}$Be and  $^{19}$C Coulomb dissociation
experiments \cite{nak94,nak99}. 

On the other hand, in the first-order semiclassical Coulomb excitation 
theory which was widely applied in the recent years to the Coulomb dissociation
of the neutron halo nuclei (see, e.g., \cite{BaurHT01}),
these effects were found to be small, for both zero range as well as
finite range wave functions of the $c+n$ system. 

One can  establish a  relation between the apparently very different
CWBA and the semiclassical theory. It was recently noticed \cite{bht01}
that in the limit of Coulomb parameter $\eta_a \ll 1$ (i.e. in the Born
approximation), both theories give the same result. It was further found
that this agreement is also valid for arbitrary values of $\eta_a$ and
$\eta_c$, provided the beam energies are high as compared 
to the relative energy ($E_{cn}$) of fragments $c$ and $n$ in the ground
state of the projectile. The first order approximation to
the amplitude given by Eq.~(11), can be written as \cite{bht01} 
\begin{eqnarray}
{\hat \ell}\beta^{\textrm {first order}} _{\ell m} & = & 4\pi Z_\ell 
f_{\textrm{coul}} e^{-\frac{\pi}{2}\xi}\nonumber \\ \times
&&\left[e^{-i\phi}\frac{1}{k_a^2 - \left[\vec k_c + \delta\vec k_n \right]^2}
+ e^{i\phi}\frac{m_c}{m_a}\frac{1}{\left[k_c^2 - \left(\delta\vec k_n -
\vec k_a\right)^2 \right]} \right],
\end{eqnarray} 
where the relative phase $\phi = \sigma(\eta_c) - \sigma(\eta_a) -
\sigma(\xi) - \xi/[2\log|D(0)|]$, with $\sigma(\eta)$ being the usual
Coulomb phase shifts, and $\xi = \eta_c - \eta_a$. In Eq. (23), we have 
defined $f_{\textrm {coul}}= 2\eta_a k_a/k^2$. This term is very similar 
to the Born approximation (BA) result given in \cite{BaurHT01};
in the limit $\xi\rightarrow 0$ it actually coincides with the BA expression. 
This equation can be used to investigate the role of higher order effects. 
\subsection{Scaling Properties}
\label{sec:2}

In many experimental situations, the momentum transfer $\vec k$
[Eq.~(20)] is small. One can expand  Eq.~(23)
(with $\phi = \xi = 0$) for small values of $k$ to obtain
\begin{eqnarray}
{\hat \ell}\beta_{\ell m}^{\textrm {first order}} &=& f_{\textrm{coul}}
\frac{2 Z_\ell}{\pi^2} 
\frac{m_n^2 m_c}{\left(m_n+m_c\right)^3}
\frac{2 \vec q \cdot \vec k}{\left(\kappa^2+q^2\right)^2},
\end{eqnarray}
where the relative momentum between c and n is given by
$\vec q=\frac{m_c \vec k_n - m_n \vec k_c}{m_n+m_c}$, and $\kappa$ is
related to the $c-n$ separation energy in the ground state of the projectile 
$E_{cn}^{bind} (= \hbar^2\kappa^2/2\mu$, with $\mu$ being the reduced mass
of the $c$-$n$ system). This result is in remarkable agreement with the usual
first order treatment of the electromagnetic excitation in the semiclassical
approximation.

In the semiclassical approach, the scattering amplitude is given
by the elastic scattering (Rutherford) amplitude times an 
excitation amplitude $a(b)$, where the impact parameter $b$ is related
to the transverse momentum transfer, $q_\perp$, and $\eta_a$ 
by $b = 2\eta_a \hbar/q_\perp$.  The absolute square of $a(b)$ gives
the breakup probability $P(b)$, which, in the lowest order (LO),
is given by \cite{BaurHT01,TypelB01} 
\begin{eqnarray}
\frac{dP_{\textrm{LO}}}{dq} = \frac{16 y^2}{3\pi\kappa} \frac{x^4}{(1+x^2)^4},
\end{eqnarray}
where variable $x$ is related to the relative momentum between $n$ and $c$
by $x=\frac{q}{\kappa}$ and $y$ is a strength parameter given by
\begin{equation}
y = \frac{2 Z_tZ_c m_n e^2}{\hbar v_a (m_c+m_n) b \kappa}.
\end{equation} 
This formula shows very interesting scaling properties: very many
experiments, for neutron halo nuclei with different binding energy,
beam energy, and scattering angles (or $\vec q_n$ and $\vec q_c$) {\em all
lie on the same universal curve}!  The deviations from this simple
scaling behavior, e.g., postacceleration effects, will lead to
violations of such scaling.

\section{Numerical Results}
We now investigate the breakup of the one-neutron halo nuclei $^{11}$Be
and $^{19}$C.  We take a heavy target of atomic number Z=82. In this 
paper, all the higher order results correspond to calculations 
performed within the finite range CWBA model, while the first order
results have been obtained by using Eq.~(23). The structure term
$Z_\ell^\prime$ [Eq.~(21)] was calculated by adopting a single particle
potential model to obtain the ground state wave function of the projectile.
The ground state of $^{11}$Be was assumed to have a 2$s_{1/2}$ valence
neutron coupled to the $^{10}$Be$(0^+)$ core with a binding energy of
504 keV. The corresponding single particle wave function was constructed by
assuming the neutron - $^{10}$Be interaction of the Woods - Saxon type.
For a set of values of the radius and diffuseness parameters (same as those
given in \cite{cha00}), the depth of this potential was searched so as to 
reproduce the ground state binding energy. This 2$s_{1/2}$ wave function has
an additional node as compared to a simpler zero-range wave function.
For the $^{19}$C, the ground state was assumed to  have a  
[$^{18}$C(0$^+$) $\otimes 2s_{1/2}\nu$] configuration with a separation
energy of 0.530 MeV. The radius and diffuseness parameters, used in the
well-depth search, were the same as those given in \cite{cha00}.

In Fig.~1, we present calculations for the triple differential cross sections
for the breakup reaction $^{11}$Be + Pb $\rightarrow$ n + $^{10}$Be + Pb,
as a function of the energy of the $^{10}$Be core (E$_c$),
for four beam energies
lying in the range of 5 MeV/nucleon - 72 MeV/nucleon. To see the
postacceleration in a clear way, it is very useful to study the cross-section
as a function of the core energy. The results obtained within the higher order
and the first order theories are shown by solid and dotted lines, respectively.

It can be seen from this figure that while for lower beam energies, the
higher order and first-order results differ considerably from each other,
they are almost the same for the beam energy of 72 MeV/nucleon. In each case,
the first order cross sections peak at the energy of the core fragment which
corresponds to the beam velocity (this value of the core fragment energy
will be referred to as E$_{bv}$ in the following). In contrast to this,
the peaks of the
higher order cross sections are shifted to energies $>$ E$_{bv}$ for the
three lower energies. Only for the 72 MeV/nucleon beam energy, does the
higher order result peak at E$_{bv}$. This shows very clearly that the
finite range CWBA model exhibits postacceleration for
beam energies $\leq$ 30 MeV/nucleon, while this effect is not present at
72 MeV/nucleon. Therefore, the higher order effects are minimal for the
Coulomb breakup of $^{11}$Be at the beam energies $\geq$ 70 MeV. This 
result is in agreement with those obtained in \cite{TypelB01,ts01}.
\begin{figure}
\begin{center}
\mbox{\epsfig{file=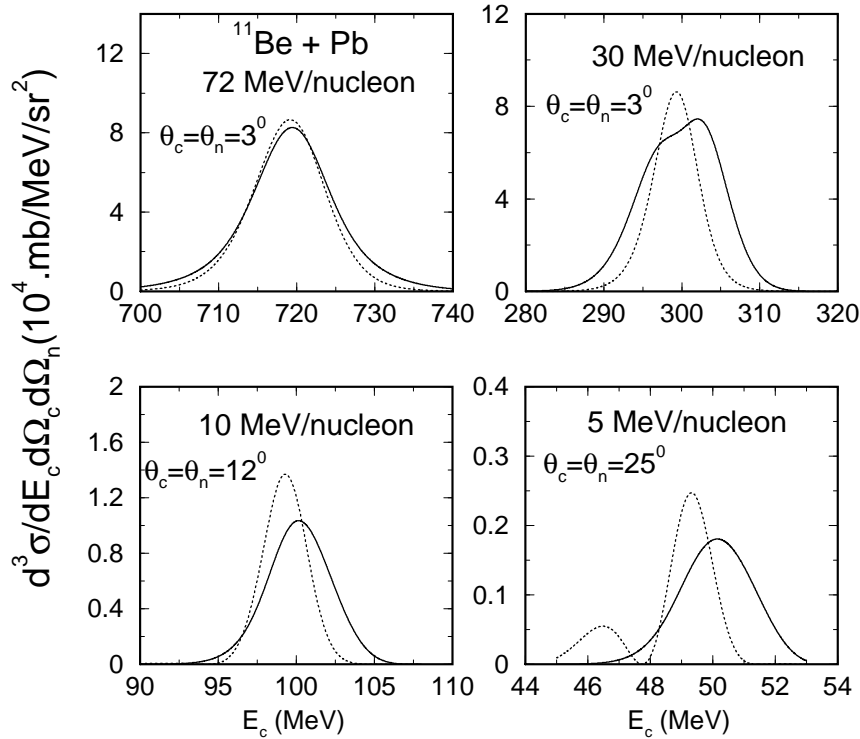,height=10.0cm}}
\end{center}
\caption {
Triple differential cross section as a function of the energy of 
$^{10}$Be core for the reaction $^{11}$Be + Pb $\rightarrow$ n + $^{10}$Be
+ Pb at the beam energies of 72 MeV/nucleon, 30 MeV/nucleon, 10 MeV/nucleon
and 5 MeV/nucleon. The results of the finite range CWBA and first-order
theory are shown by solid and dotted lines respectively.
}
\label{fig:figa}
\end{figure}
\noindent

In Fig.~2, we compare the first-order (dotted lines) and the higher order
(solid lines) results for the double differential cross section for the
same reaction, and for the same beam energies as in Fig.~1. These results
have been obtained by integrating the triple differential cross sections
over the unobserved neutron solid angles. The cross sections are shown as
a function of E$_c$. It is clear that for beam energies
$\leq$ 30 MeV/nucleon, the first order results peak at E$_{bv}$, but the
higher order cross sections have their maxima at energies larger than
E$_{bv}$. In contrast to this, both the higher order and the first
order cross sections peak at the same value of E$_c$ (= E$_{bv}$),  
for the 72 MeV/nucleon case. Therefore, the  postacceleration effects
are pronounced for the smaller beam energies, whereas they become quite small
for the higher energies.  
The near equality of the first-order and the finite range CWBA cross sections,
at the beam energy of 72 MeV/nucleon, suggest that for this reaction,  
the higher order effects, in general, are quite irrelevant at beam
energies $\geq$ 70 MeV/nucleon.
\begin{figure}
\begin{center}
\mbox{\epsfig{file=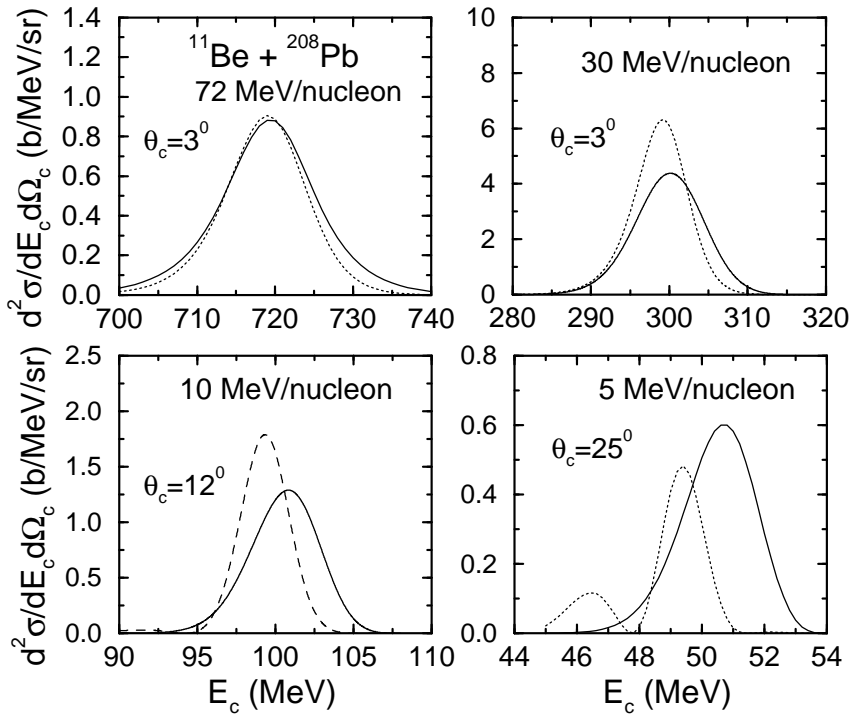,height=10.0cm}}
\end{center}
\caption {
Double differential cross section as a function of the energy of 
$^{10}$Be core for the reaction $^{11}$Be + Pb $\rightarrow$ n + $^{10}$Be
+ Pb at the beam energies of 72 MeV/nucleon, 30 MeV/nucleon, 10 MeV/nucleon
and 5 MeV/nucleon. The results of the finite range CWBA and first-order
theory are shown by solid and dotted lines respectively.
}
\label{fig:figb}
\end{figure}
In Fig.\ 3, we compare the results of the first-order and the finite range 
CWBA calculations
for the relative energy spectrum of the fragments emitted in the breakup
reaction of $^{11}$Be on a $^{208}$Pb target for the same four beam energies
as shown in Fig.\ 2. These cross sections have been obtained by integrating
over all the allowed values of the angles $\Omega_{c-n}$. In both the models, 
the integrations over $\theta_{t-(c+n)}$, have been carried out between
1$^\circ$ to grazing angle, in the upper two figures, and between
5$^\circ$ to grazing angles, in the lower two figures. The integrations
over $\phi_{t-(c+n)}$ angles have been done over all of its kinematically
allowed values. The dotted and solid lines represent the results of the
first-order and the higher order
\begin{figure}[here]
\begin{center}
\mbox{\epsfig{file=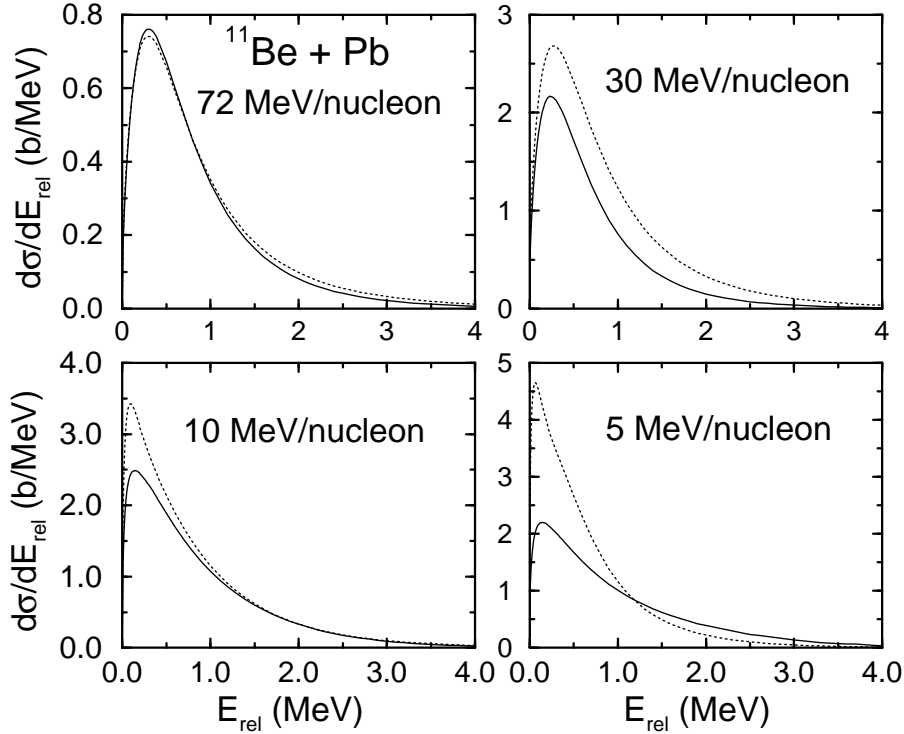,height=10.0cm}}
\end{center}
\caption {
The differential cross section as a function of the relative energy of the
fragments (neutron and $^{10}$Be) emitted in the $^{11}$Be induced
breakup reaction on a $^{208}$Pb target at the beam energies of
72 MeV/nucleon, 30 MeV/nucleon, 10 MeV/nucleon, and 5 MeV/nucleon.
The dotted and full lines represent the first-order and the finite range 
CWBA results, respectively.  
}
\label{fig:figc}
\end{figure}
\noindent
calculations, respectively. We notice that while for the beam
energy of 72 MeV/nucleon, the higher order effects are
minimal, they are quite strong  
for the lower beam energies, being largest at the beam energy of
5 MeV/nucleon. This reinforces the point, already made 
in \cite{TypelB01,ts01}, that at the beam energy of 72 MeV/nucleon, the
higher order effects are quite small if both the first order and the higher
order terms are calculated within the same theory.

In Fig.\ 4, we show the same results as in Fig. 3, but for the  
$^{19}$C induced reaction on the $^{208}$Pb target for the beam 
energies of 67 MeV/nucleon, 30 MeV/nucleon and 10 MeV/nucleon. Like
Fig. 4, the integrations over the fragments center of mass angles is 
done in the range of 1$^\circ$ to grazing angle, for first two beam 
energies and between 5$^\circ$ to grazing angle, at the lowest beam energy.
We see that in this case too the higher order effects are quite weak
for the beam energy 67 MeV/nucleon, but appreciable for the lower
beam energies.

It may be noted that by comparing the result of a conceptually different model 
of the Coulomb breakup reactions \cite{TRJ98} than ours, with that of the
first order semiclassical perturbation theory of the Coulomb excitation,
it has been concluded in \cite{tos00} that the higher order effects are
substantial for these reactions even at the beam energies $\sim$
70 MeV/nucleon. However, one should be careful in drawing definite
conclusions about the role of the higher order effects from such an
approach.     
For a reliable assessment of the contributions of the higher order
effects, it is essential that both the first order and the
higher order terms should be  calculated within the same model, as has
been done in this work (and also in \cite{TypelB01,ts01}).
\begin{figure}[here]
\begin{center}
\mbox{\epsfig{file=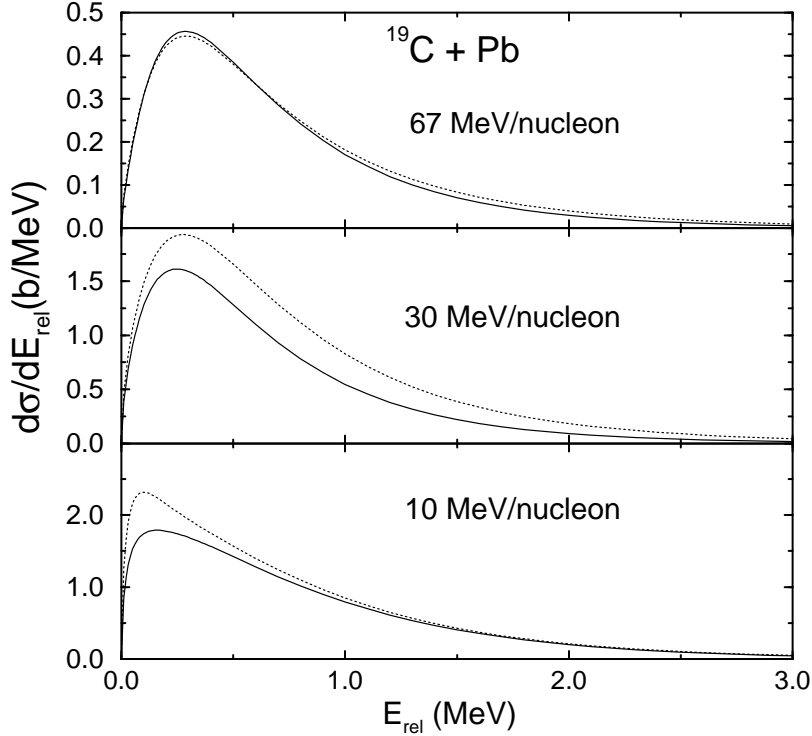,height=10.0cm}}
\end{center}
\caption{
The differential cross section as a function of the relative energy of the
fragments (neutron and $^{18}$C) emitted in the $^{19}$C induced
breakup reaction on a $^{208}$Pb target at the beam energies of
67 MeV/nucleon, 30 MeV/nucleon, and 10 MeV/nucleon.
The dotted and full lines represent the first-order and the finite range 
CWBA results, respectively.  
}
\label{fig:figd}
\end{figure}
\noindent
\section{Conclusion and  Outlook}
\label{sec:3}
In this paper, we investigated the breakup of the one-neutron
halo nuclei, $^{11}$Be and $^{19}$C, in the Coulomb field of a
heavy target nucleus, within a theory which is formulated in the framework
of the post form distorted wave Born approximation. This theory contains
the electromagnetic interaction between the core and the target nucleus to
all orders. An attractive feature of this formulation is that the
corresponding pure Coulomb breakup amplitude can be expressed in an
analytic form. We have also extracted the first order approximation of
the full pure Coulomb breakup $T$-matrix. This enables us to investigate 
the respective roles of the first order and the higher order effects within
the same theoretical model. We studied the beam energy dependence of
the first order and higher order triple and double differential 
cross sections. We also investigated the contributions of the 
higher order effects to the relative energy distribution of the fragments.  

In the higher order model, the peaks in
the triple and double differential cross sections vs. core energy spectra,
are shifted to energies larger than those corresponding to the beam
velocity, at the incident energies $\leq$ 30 MeV/nucleon. Therefore,
postacceleration effects are important at these beam energies. On the
other hand, at the beam energy $\sim$ 70 MeV/nucleon, the corresponding
spectra peak at the beam velocity energies, which is consistent with
no postacceleration. In contrast to this, the first-order cross sections
always peak at the beam velocity energy, which is expected as the
postacceleration is a higher order effect. 

The higher order effects are also found to be quite important in the
relative energy spectrum of the fragments at beam energies $\leq$ 30 
MeV/nucleon, while they are insignificant at the beam energies $\sim$
70 MeV/nucleon. This suggests that the conclusions arrived at in Refs.
\cite{nak94,nak99}, where the data on the relative energy spectra of
the fragments taken in the breakup of $^{11}$Be and $^{19}$C at 
beam energies $\sim$ 70 MeV, have
been analyzed within the first order theory of the Coulomb excitation, 
may not be affected by the higher order effects. 
 
The present model can be seen as a ``theoretical laboratory'', which allows to
study  numerically the relation between quantal and
semiclassical theories, and the importance of postacceleration effects.
It should be noted that from an experimental point of view, the
postacceleration
effects are not fully clarified (see, e.g., \cite{nak94,Ieki93,Bush98}).
Finally, let us mention the recent work on the electromagnetic dissociation
of unstable neutron-rich oxygen isotopes \cite{Leistenschneider01}.
These authors  deduce photoneutron cross-sections 
from their dissociation measurements. 
If the neutrons are emitted in a slow evaporation process in a later stage of 
the reaction, the question of postacceleration is not there. On the other hand,
for the light nuclei there is some direct neutron emission component and
the present kind of theoretical analysis further  proves the validity
of the semiclassical approach used in \cite{Leistenschneider01}.

Postacceleration effects are also of importance for the use of Coulomb 
dissociation for the study of radiative capture reactions of
astrophysical interest. We expect that our present investigations will
shed light on  questions of  postacceleration and 
higher order effects in these cases also.

\end{document}